\documentstyle[referee]{laa}

\begin{document}
\thesaurus{07  
	   (07.09.1; 
	    07.13.1;  
	   )}

\title{Solar Radiation and Asteroidal Motion \\
\vspace*{1.cm}
{\it Letter to the Editor}}
\author{J.~Kla\v{c}ka}
\institute{Institute of Astronomy,
   Faculty for Mathematics and Physics, Comenius University \\
   Mlynsk\'{a} dolina, 842~48 Bratislava, Slovak Republic}
\date{}
\maketitle

\begin{abstract}
Effects of solar wind and solar electromagnetic radiation on
motion of asteroids are discussed. The results complete the statements
presented in Vokrouhlick\'{y} and Milani (2000).

As for the effect of electromagnetic radiation, the complete equation of motion
is presented to the first order in $v/c$ -- the shape of asteroid
(spherical body is explicitly presented)
and surface distribution of albedo should be taken into account. Optical
quantities must be calculated in proper frame of reference.


\end{abstract}

\section{Introduction}
Vokrouhlick\'{y} and Milani (2000) investigate motion of asteroids due to
direct solar radiation pressure. They have come to the conclusion that
Poynting-Robertson effect (P-R effect; Robertson 1937)
should be also considered.

The aim of this paper is to call attention towards
i) the significance of the solar wind in comparison with the P-R effect,
and, ii) the possible significance of the real equation of motion
originating from the interaction of solar electromagnetic radiation
with an asteroid.

\section{Solar Wind}
As for the effect of the solar wind, we use the equation of motion
derived in Kla\v{c}ka (1994), see also Kla\v{c}ka (1999a).
As a consequence, application to 1566 Icarus, the effect of solar wind
yields, e. g., that the current secular decrease of semi-major axis
is in 0.20 larger than for the P-R effect for perfect absorption;
the value 0.20 is a little larger than the value 4$a_{0}$/9 used in
Vokrouhlick\'{y} and Milani (2000).

\section{Solar Electromagnetic Radiation}
Taking into account Eqs. (6), (7) and (29) in Kla\v{c}ka (2000c)
we can write equation of motion for spherical
body due to its interaction with electromagnetic radiation:
\begin{eqnarray}\label{1}
\frac{d \vec{v}}{d t} &=& \frac{S}{m~c} ~ \pi ~R^{2} ~
	\left \{ \left ( 1 ~-~ \frac{\vec{v} \cdot \hat{\vec{S}}_{i}}{c} \right ) ~
	\hat{\vec{S}}_{i} ~-~\frac{\vec{v}}{c} \right \} ~+~
	\frac{S}{m~c} ~ \frac{2}{3} ~R^{2} ~ X ~~,
\nonumber \\
 X &=&	\left \{ \left ( 1 ~-~ \frac{\vec{v} \cdot \hat{\vec{S}}_{i}}{c} \right ) ~
	\hat{\vec{S}}_{i} ~-~\frac{\vec{v}}{c} \right \} ~
	\int_{0}^{2~\pi} d~ \varphi ' ~ \int_{0}^{\pi / 2} d~ \vartheta ' ~
	a'_{0} ( \vartheta ', \varphi ') ~ sin \vartheta ' ~
	\cos ^{2} \vartheta ' ~-~
\nonumber \\
  & & -~ \left \{ \left ( 1 ~-~ 2~ \frac{\vec{v} \cdot \hat{\vec{S}}_{i}}{c}
	 ~+~ \frac{\vec{v} \cdot \hat{\vec{e}}_{1}}{c}	\right ) ~
	\hat{\vec{e}}_{1} ~-~ \frac{\vec{v}}{c} \right \} \times
\nonumber \\
 & & \times  \int_{0}^{2~\pi} d~ \varphi ' ~ \int_{0}^{\pi / 2} d~ \vartheta ' ~
	a'_{0} ( \vartheta ', \varphi ') ~ sin ^{2} \vartheta ' ~
	\cos \vartheta ' ~ \cos \varphi ' ~-~
\nonumber \\
  & & -~ \left \{ \left ( 1 ~-~ 2~ \frac{\vec{v} \cdot \hat{\vec{S}}_{i}}{c}
	 ~+~ \frac{\vec{v} \cdot \hat{\vec{e}}_{2}}{c}	\right ) ~
	\hat{\vec{e}}_{2} ~-~ \frac{\vec{v}}{c} \right \} \times
\nonumber \\
  & &	\int_{0}^{2~\pi} d~ \varphi ' ~ \int_{0}^{\pi / 2} d~ \vartheta ' ~
	a'_{0} ( \vartheta ', \varphi ') ~ sin ^{2} \vartheta ' ~
	\cos \vartheta ' ~ \sin \varphi ' ~,
\end{eqnarray}
where $a'_{0} ( \vartheta ', \varphi ')$ is surface albedo of a spherical body
(asteroid) of mass $m$ and radius $R$ moving around the Sun with orbital
velocity $\vec{v}$, $c$ is the velocity of light,
$S$ is the flux density of the solar electromagnetic radiation,
$\hat{\vec{S}}_{i}$ is unit vector of the incident radiation,
unit vectors
$\hat{\vec{e}}_{1} = ( 1 ~-~ \vec{v} \cdot \hat{\vec{e}}_{1} ' / c )
		    ~ \hat{\vec{e}}_{1} ' ~+~ \vec{v} / c$,
$\hat{\vec{e}}_{2} = ( 1 ~-~ \vec{v} \cdot \hat{\vec{e}}_{2} ' / c )
		    ~ \hat{\vec{e}}_{2} ' ~+~ \vec{v} / c$
(analogous relation holds for $\hat{\vec{S}}_{i}$),
$\hat{\vec{e}}_{1} ' \cdot \hat{\vec{e}}_{2} ' =$ 0,
$\hat{\vec{e}}_{1} ' \times \hat{\vec{e}}_{2} ' = \hat{\vec{S}}_{i} '$.
$\varphi ' = 0$ for the direction and orientation $\hat{\vec{e}}_{1} '$,
$\vartheta '$ is measured from $-~\hat{\vec{S}}'_{i}$,
both angles
are measured in positive directions.
The numerical factor 2/3 comes from
the considered model -- diffuse reflection corresponding to the
Lambert's law (although it may not hold for real bodies in Solar System --
see p. 112 in Van de Hulst (1957)):
$\int_{2~\pi}$ $\left \{ \left ( \cos \alpha \right ) / \pi
\right \} ~ \cos \alpha ~ d \Omega = 2/3$,
$d \Omega = 2~ \pi ~ \sin \alpha ~ d \alpha$.

If albedo $a'_{0}$ is a constant for the whole sphere, then Eq. (1)
reduces to
\begin{eqnarray}\label{2}
\frac{d \vec{v}}{d t} &=& \frac{S}{m~c} ~ \pi ~R^{2} ~ \left ( 1 ~+~
	\frac{4}{9} ~a'_{0}  \right ) ~
	\left \{ \left ( 1 ~-~ \frac{\vec{v} \cdot \hat{\vec{S}}_{i}}{c} \right ) ~
	\hat{\vec{S}}_{i} ~-~\frac{\vec{v}}{c} \right \} ~,
\end{eqnarray}
which is the P-R effect to the first order in $v/c$ for our case
($Q'_{PR} = 1 ~+~ 4 a'_{0} / 9$ -- see Eq. (122) in Kla\v{c}ka 1992a).

\subsection{Solar Electromagnetic Radiation -- Discussion}
Eq. (1) represents equation of motion for interaction between spherical
particle and electromagnetic radiation (thermal reemission of the absorbed
energy connected with thermal conductivity of the surface material
is not considered). As it was already stressed a special
form of Eq. (1) is Eq. (2) which is the P-R effect.

Eq. (1) is real relativistic equation of motion to the first order in $v/c$.
The access of Vokrouhlick\'{y} and Milani (2000) corresponds to putting together
some of its individual terms in a heuristic manner. They seem to be in
coincidence with Eq. (1) if we neglect less significant terms in Eq. (1).
However, our access is physically
more correct since it shows that no other terms are important; moreover,
some new terms of the order of $\vec{v} / c$ are presented. The important
fact which one must bear in mind is that values of integrals in Eq. (1)
correspond to proper (local) inertial frame of reference.

P-R effect is not the acceleration which is presented in
Vokrouhlick\'{y} and Milani (2000), as it is immediately seen from
the comparison of our Eq. (2) and their Eq. (21). As for the correct
equation for secular changes of orbital elements, together with
initial conditions, we refer the reader
to Kla\v{c}ka (1992b) and Kla\v{c}ka and Kaufmannov\'{a} (1992).
Vokrouhlick\'{y} and Milani (2000) refer to Breiter and Jackson (1998).
However, except confirming the results of Wyatt and Whipple (1950),
(see Kla\v{c}ka (1992b), for more details; initial conditions are also
important) and Kla\v{c}ka and Kaufmannov\'{a} (1992) the paper
by Breiter and Jackson (1998) yields only nonphysical results
(nice analytical solution yielding still increasing eccentricity does not occur
in reality due to higher orders in $v/c$ in the P-R effect), as it
is presented in Kla\v{c}ka (1999b).
As for correct understanding of the P-R effect,
we refer also to Kla\v{c}ka (2000b). As an soluble example for secular
changes of orbital elements for nonspherical particle we refer to
Kla\v{c}ka (2000a).
As for relativistic covariant equation of motion we refer to Kla\v{c}ka (2000c).

\section{Conclusion}
We have shown the way in which the effect of solar wind must be taken into
account in dealing with the motion of a body (asteroid).

We have presented complete equation of motion (to the first order in
$\vec{v} / c$) for spherical body
under interaction with the electromagnetic radiation.
The important statement concerns the fact that optical quantities
must be calculated in the proper frame of reference of the body.

We have called attention towards the correct physics as for the P-R effect
and solution of the corresponding equation of motion.

\acknowledgements
Special thanks to the firm ``Pr\'{\i}strojov\'{a} technika, spol. s r. o.''.
The paper was partially
supported by the Scientific Grant Agency VEGA (grant No. 1/7067/20).

\end{document}